# Electronic structure and correlations in planar trilayer nickelate Pr$_4$Ni$_3$O$_8$

Haoxiang Li[1,2]*, Peipei Hao[1], Junjie Zhang[3,4], Kyle Gordon[1], A. Garrison Linn[1], Hong Zheng[3], Xiaoqing Zhou[1], J.F. Mitchell[3]*, D. S. Dessau[1,5]*

[1]*Department of Physics, University of Colorado Boulder, Boulder, CO 80309, USA*

[2]*Advanced Materials Thrust, Hong Kong University of Science and Technology (Guangzhou), Guangzhou, Guangdong 511458, China*

[3]*Materials Science Division, Argonne National Laboratory, Lemont, IL 60439, USA*

[4]*Institute of Crystal Materials & State Key Laboratory of Crystal Materials, Shandong University, Jinan, Shandong 250100, China*

[5]*Center for Experiments on Quantum Materials, University of Colorado Boulder, Boulder, CO 80309, USA*

*Correspondence and requests for materials should be addressed to D.S.D. (dessau@colorado.edu), J.F.M(mitchell@anl.gov), H.L. (haoxiangli@ust.hk)

**The recent discovery of superconductivity in hole-doped planar nickelates R$_{1-x}$SrNiO$_2$ (R=Pr,Nd) raises the foundational question of how the electronic structure and electronic correlations of these Ni$^{1+}$ compounds compare to those of the Cu$^{2+}$ cuprate superconductors. Here, we present an Angle-Resolved Photoemission Spectroscopy (ARPES) study of the trilayer nickelate Pr$_4$Ni$_3$O$_8$, revealing an electronic structure and Fermi surface very similar to that of the hole-doped cuprates but with a few critical differences. Specifically, the main portions of the Fermi surface are extremely similar to that of the bilayer cuprates, with an additional piece that can accommodate additional hole doping. We find that the electronic correlations are about twice as strong in the nickelates and are almost k-independent, indicating that they originate from a local effect – likely the Mott interaction, whereas the cuprate interactions are somewhat less local. Nevertheless, the nickelates still demonstrate an** **approximately linear in energy and linear in temperature scattering rate. Understanding the similarities and differences between these two related families of strongly-correlated novel superconductors is an important challenge.**

## Introduction

As $3d$ transition metal oxides, various nickelate compounds have long been studied for correlated electron properties and as candidates for high-$T_c$ superconductivity. In particular, special attention has been paid to two-dimensional $Ni^{1+}$ compounds that might mimic the $Cu^{2+}$ state of the cuprates that is known to support both the strange-metal physics and high $T_c$ superconductivity. However, even with the same formal electron count between the $Ni^{1+}$ and $Cu^{2+}$ states, concerns have been raised that these two states would not be the same and so could not support the same exotic physics as the cuprates, with concerns arising due to reduced $3d$-$2p$ mixing and different on-site correlation energies [1]. Such differences could arise due to the varying energetics of the average transition metal $d$ bands as a function of atomic number, as described in the Zaanen-Sawatzky-Allen classification schemes of correlation gaps as arising from Mott correlations that principally center on the transition metal (farther left in the periodic table) or charge-transfer physics (farther right) [2] that includes the oxygen ligands more explicitly and brings in the Zhang-Rice singlet that is a combination of the Cu and O orbitals [3].

Recently, the spotlight has been turned to square-planar nickelate materials following the breakthrough discovery of superconductivity in thin films of the planar infinite layer nickelates, Sr-doped $RNiO_2$ (R=Pr, Nd) [4,5], and more recently the 5-layer variant [6]. It is natural that proposals have been put forward that the superconducting nickelate should express exotic electronic correlations similar to those found in other unconventional superconductors, such as the doped charge-transfer cuprates [7,8,9,10,11]. However, the topotactic reduction process used in the synthesis of these nickelate thin films, which is required to remove the apical oxygen atoms and thus to create the square planar network, is known to degrade the surface. This has so far prohibited investigations necessary to make direct measurements of the electronic structure and test such conjectures because the most direct probes of the electronic structure and carrier dynamics (e.g., ARPES) require surface-sensitive techniques.

We focus on the planar trilayer nickelates $Pr_4Ni_3O_8$ that can be grown in bulk, cleavable crystalline form suitable for ARPES [12]. In this material, the apical oxygen has been removed from the Ruddlesden-Popper (R-P) phase $Pr_4Ni_3O_{10}$ structure by reduction, leading to square planar nickel assembled into triple $NiO_2$ layers that alternate with fluorite-like $R_2O_2$ layers (Fig. 1A). In contrast to the R-P nickelates such as $RNiO_3$, $R_2NiO_4$, and $R_4Ni_3O_{10}$ that several previous studies have focused on [13,14,15,16,17,18], the planar nickelates–e.g., $R_{1-x}Sr_xNiO_2$, $R_4Ni_3O_8$, are proximate to the $d^9$ electron count per transition metal ion that is common to the cuprates. Based on the electron counting of the $3d$ orbital states, the planar trilayer nickelates possess a 1/3-hole doping with average orbital occupancy $d^{8.67}$, which nominally is on the heavily overdoped side of the high $T_C$ cuprate phase diagram (Fig. 1B). Furthermore, a recent study has revealed a large orbital polarization of $dx^2$-$y^2$ character in the $3d$ conduction bands of $R_4Ni_3O_8$—a key signature of the cuprate electronic structure [19]. Nevertheless, no experimental measurement of the electronic structure and carrier dynamics of these materials has yet been reported.

**Results**

We performed high resolution ARPES measurements on single crystals of trilayer nickelate $Pr_4Ni_3O_8$, cleaved in-situ in the ultra-high vacuum environment of the ARPES spectrometer. To explore the electronic structure, we first present the Fermi surface measured at T=22K in Fig. 1C,D. By using different photon energies (84eV and 55eV), we take advantage of the matrix element effects to highlight distinct parts of the Fermi surface. The Fermi surface maps clearly display an electron pocket centered at $\Gamma$ (first zone) and $\Gamma'$ (second zone) (Fig. 1C), and a hole pocket or pockets centered around the zone corner (Fig. 1D). $Pr_4Ni_3O_8$ has a much simpler Fermi surface than that of the related trilayer Ruddlesden-Popper nickelate $La_4Ni_3O_{10}$, which has a significantly different electron count, $d^{7.33}$ [13], and an additional piece of Fermi surface near the $\Gamma$ point not found in either the cuprates or the present compound. In Fig. 1E, we sketch the two main Fermi pockets from panels C&D in solid line, while including a third pocket in dashed line that is more difficult to resolve from those experimental panels. We compare these results with the Fermi surface of optimally doped bilayer cuprate $Bi_2Sr_2CaCu_2O_{8+x}$ (Bi2212) (Fig.1F). We see that the Fermi surface topologies are extremely similar, except for the additional inner piece of Fermi surface of $Pr_4Ni_3O_8$ that accommodates the extra doping of holes in that material. The DFT result further supports our findings from ARPES (Fig. 1G). However, in order to avoid issues related to the $4f$ state of Pr in the DFT calculation that contradicts the experiment results [19] (see Materials and Methods, and Supplementary Text S7), we perform the DFT calculations on the closely related compound $La_4Ni_3O_8$ instead, an approach taken in another DFT calculations on this material [20] and on the infinite layer nickelate [21]. Here we note that both La and Pr are expected to be in the 3+ valence state, and so the valence of Ni and O in $Pr_4Ni_3O_8$ and $La_4Ni_3O_8$ are expected to be extremely similar—something that is confirmed by the excellent agreement between the experimental and theoretical Fermi surfaces shown here. It is worth noting that recent DFT works on $Pr_4Ni_3O_8$ applied the Coulomb repulsion values $U$ in different orbital states to purposely remove the contamination from the Pr $4f$ states, which return an almost identical $3d$ band structure near the Fermi level compared to our result from $La_4Ni_3O_8$ [22,23] (see Fig. S6 for the comparison of the DFT results).

Figure 2 presents another aspect of the electronic structure, in which we show the ARPES dispersions along the high symmetry cuts. Figs. 2A,B denote the position of the high symmetry cuts, while the solid (dashed) curves indicate the parts of the Fermi surface that are emphasized (suppressed) by the photon energy choices. The red and blue dots plotted in Fig. 2C-E are the spectral peak positions extracted from momentum distribution curves (MDCs) and energy distribution curves (EDCs), respectively. These spectral peak positions quantify the dispersion in the spectra. Cuts 1 and 2 represent the high symmetry cuts taken through the electron pocket shown in Fig. 2A, in which we observe sharp dispersion down to ~0.1eV below the Fermi level. In the panels to the right of each spectrum, extracted dispersions are compared to the DFT bands. The chemical potential of the DFT result has been shifted up ~10meV to match the experimental result, but without any other modifications made. This small chemical potential shift points to an electron count that is slightly away from $d^{8.67}$ (or 33% hole doping), which is likely due to the imperfect stoichiometry of the sample. With this simple rigid shifting, both cuts 1 and 2 display an excellent match to the Fermi momentum ($k_F$) of the DFT band. The X-M cut in Fig. 2E is along the so-called "antinodal cut" in hole doped cuprates, which shows a single band with the shallow band bottom at ~25meV, qualitatively similar to that in the hole doped cuprates, where the bottom of the flat bands range from ~ 100 meV below $E_F$ to just above $E_F$ (for the antibonding band of the heaviest hole doped cuprates) [24]. In the DFT calculations, there are two distinct DFT bands crossing the Fermi level along the X-M direction, while our ARPES measurement only identifies one band and it matches well in $\mathbf{k}_F$ to the DFT band that disperses down to higher binding energy. The extra DFT band in Fig. 2E is part of the invisible hole pocket that we mentioned in the context of Fig. 1E. In the calculations, this extra band predominantly originates from the outer Ni-O plane in the trilayer structure, whereas the two Fermi pockets that match the ARPES observations have a mixed weighting from both the inner and the outer Ni-O plane (see Supplementary Materials Fig. S7). One possible explanation of this discrepancy between DFT and the ARPES observation is the matrix element effect. The matrix element depends on a combination of photon energies, polarizations, experimental geometries, and the sample cleavage plane, all of which can influence the spectral intensity on certain parts of the electronic structure. The total parameter space of the matrix element is hard to exhaust; we show a photon energy scan along the X-M cut (cut 3) in the Supplementary Materials Fig. S3 that explore the matrix element effect, but we found no clear evidence of an extra splitting of the hole band. Another possibility is that the actual band splitting is much weaker than predicted by DFT, and the actual splitting is less than our instrumental resolution.

The polarization dependence of our ARPES results (part of the matrix element effects discussed above) indicates that the states which cross the Fermi energy in $Pr_4Ni_3O_8$ originate from predominantly $dx^2\text{-}y^2$ symmetry orbitals – a fact that is confirmed by our DFT calculations (see Supplementary Text S1 and Fig. S1 for both the ARPES and DFT results). This finding is consistent with polarization-dependent X-ray absorption measurements reported on these materials [19]. The dominant $dx^2\text{-}y^2$ orbital character of $Pr_4Ni_3O_8$ is a further critical aspect of the electronic structure analogy between these compounds and the cuprates.

All the DFT bands shown in Fig. 2 disperse to much higher binding energy compared to the ARPES dispersion, indicating a strong mass enhancement effect. We quantify this finding in Fig. 3. Figures. 3A-C shows the experimental dispersion (red and blue dots) from Fig. 2, compared to renormalized DFT band dispersions (black dashed lines), where the effective masses of the DFT bands have been renormalized for best agreement to the experimental dispersions (scaling the DFT dispersions to match the ARPES dispersions). The result present overall excellent agreement between the measured and renormalized DFT bands. We note that the DFT renormalization was done while keeping the chemical potential constant at the shifted value, i.e. the renormalization is centered around $E_F$ as expected for a many-body electronic interaction.

Fig. 3D shows the renormalization factors used for the DFT bands in panels A-C, i.e. the effective mass ratio of the ARPES dispersion relative to the corresponding DFT band mass, indicating a strong mass enhancement with values between 4 and 5 for all parts of the Fermi surface (the renormalization factors extracted from Fermi velocities return almost identical results, see Fig. S4 and discussions in Supplementary Materials). These values are approximately a factor of two larger than that of $La_4Ni_3O_{10}$ [13], and the strange metal state of cuprates, as will be discussed later. This strikingly large mass enhancement is a signature of a strong electron correlation effect, which is unexpected if, as implied by its nominal hole count, $Pr_4Ni_3O_8$ can be considered analogous to the heavily hole-doped cuprates. In the cuprates, such heavily doped materials are found to have reduced correlation effects compared to lower-doped compositions that lie nearer to the Mott insulating, strongly-correlated regime of the phase diagram (Fig. 1B) [25]. In addition, the large mass enhancement in $Pr_4Ni_3O_8$ is not accompanied by a kink anomaly, such as that found in the superconducting state of cuprates [26,27], as Fig. 3A-C display smooth dispersions without any obvious kink features. Because the kink anomaly is a signature of the superconducting state in cuprates [26,27], the absence of such a kink anomaly potentially reflects the missing superconductivity in $Pr_4Ni_3O_8$. However, the literature is currently silent about the existence and/or importance of any kink features in $Nd_{1-x}Sr_xNiO_2$ films, calling for further investigations to shed light on the potential relevance of such band structure anomalies to superconductivity in the nickelate family.

The strong electron correlation in $Pr_4Ni_3O_8$, signified by the large mass enhancement, also manifests through the quasiparticle scattering rates, which are nominally proportional to the ARPES peak widths. More precisely, we can extract the imaginary part of the electron self-energy function if the spectral dispersion can be approximated by a linear function (see detailed explanation in Supplementary Text S5). In this case, the imaginary self-energy $\Sigma''(\omega)$ is given by [24,25,28]:

$$\Sigma''(\omega) = v_B \frac{\Gamma_{MDC}(\omega)}{2} \tag{1}$$

where $\Gamma_{\text{MDC}}(\omega)$ is the MDC width and $v_B$ is the bare Fermi velocity, which we obtained from the bare (unrenormalized) DFT band velocity. The orange circles in Fig. 3E show the extracted $\Sigma''(\omega)$ from the "nodal cut" ($\Gamma$-M cut or "cut 2", see Fig. S9 for comparison of the MDCs and fits) where the band dispersion is nearly linear. For all energies above $\sim 15$ meV, the extracted $\Sigma''(\omega)$ displays a linear energy dependence that deviates from the quadratic behavior of the conventional Fermi liquid model (red dashed curve) which we write in a generalized form [25] as:

$$\Sigma''_{\text{FL}}(\omega) = \lambda[(\hbar\omega)^2 + (\beta k_B T)^2]/\hbar\omega_N + \Sigma''_0 \tag{2}$$

where $T$ is the temperature, $\lambda$ defines the general coupling scale, $\beta$ is a scaling factor for $k_B T$, $\Sigma''_0$ is an offset parameter that accounts for impurity or disorder scattering and $\omega_N$ is a normalization frequency that maintains the proper dimensionality of the self-energy [25]. On the contrary, the marginal Fermi liquid model (MFL) that was used to describe the strange metal states in cuprates presents an excellent fit to the data. The MFL model was first proposed to explain the non-Fermi liquid behavior found in high $T_C$ superconductor cuprates [29], which can be written as [25]:

$$\Sigma''_{\text{MFL}}(\omega) = \lambda\sqrt{(\hbar\omega)^2 + (\beta k_B T)^2} + \Sigma''_0 \tag{3}$$

By fitting the data with the parameters $\lambda$, $\beta$ and $\Sigma''_0$ (see Supplementary Text S5 for details about the extracted parameters) we get the fit shown by the black dashed lines in Fig 3E, which is very nearly linear in energy for $\omega \gg \beta k_B T$, i.e. beyond about 15 meV (green line), whereas near $E_F$ the finite temperature leads to some curvature according to Eqn. 3. This same form of quasiparticle scattering (imaginary self-energy) is also approximately represented in the resistivity curve versus temperature, though with different proportionality constants that are in general much more complicated to determine [30,31]. Fig. 3F plots the normalized resistivity curve $\rho(T)/\rho(300 \text{ K})$ together with a quadratic fit (red dashed curve) and a linear fit (solid green line). Similar to the ARPES quasiparticle scattering rate, the linear fit agrees better with $\rho(T)$ than does the quadratic fit, except for the low temperature region (<30K), which is believed to be dominated by impurity scattering in this compound [19]. Therefore, the combined behavior of both the ARPES and resistivity data indicates scattering that is nominally linear in both $\omega$ and $T$ (Eqn. 3), rather than the more standard quadratic Fermi Liquid behavior implied by Eqn. 2. It is worth noting that the general evolution of the electron scattering from Fermi liquid to non-Fermi liquid behavior can be described by the Power Law Liquid model in ref. [25].

The linear-scattering MFL behavior found in $Pr_4Ni_3O_8$ establishes that its electron dynamics are analogous to that found in the cuprates. Nevertheless, further comparison of the imaginary self-energy in these two types of materials reveals that $Pr_4Ni_3O_8$ in fact hosts much stronger electronic correlations than the cuprates, consistent with the greater mass enhancements derived from the dispersion data. Figures 4A-D compares the $\Gamma$-M or nodal cut of $Pr_4Ni_3O_8$ (cut 2) to that from three cuprate compounds in the strange metal state above their superconducting transitions. The cuprate spectra display sharper and more-coherent peaks, especially in the optimally doped Bi2212, whereas the $Pr_4Ni_3O_8$ spectrum is dramatically broadened towards high binding energies. The $\Sigma''(\omega)$ extracted from MDC widths as in Eqn. 1 are plotted in Fig. 4E, quantifying this effect. The $\Sigma''(\omega)$ of $Pr_4Ni_3O_8$ exhibits a much faster rising slope than the cuprate data, presenting a remarkably stronger dynamic of the electronic interactions. The zero-frequency offset of $\Sigma''(\omega)$ is usually attributed to the less-interesting impurity or elastic terms. This zero-frequency offset of $\Sigma''(\omega)$ in $Pr_4Ni_3O_8$ is similar to what we measured in $La_{2-x}Sr_xCuO_4$ (LSCO - inset of Fig. 4E), which indicates that the $Pr_4Ni_3O_8$ crystal quality is comparable to the LSCO crystals. The dimensionless coupling constant $\lambda$ (Eqn. (3)), extracted from $\Sigma''(\omega)$, describes the intensity of the dynamical excitations and takes the value $\lambda > 4$ in $Pr_4Ni_3O_8$, greatly surpassing the one in cuprates ($\lambda \sim 0.5$ to 1.0) as shown in Fig. 4 [25]. To our knowledge, this is one of the largest linear energy slopes (or dynamic excitation) found in $\Sigma''(\omega)$ among a variety of well-studied strongly correlated materials, including cuprates, iridates, and ruthenates. [25, 32, 33,34].

The dynamic excitation in $\Sigma''(\omega)$ is directly connected to the real part of self-energy $\Sigma'(\omega)$ by the Kramers-Kronig relation [25,35]. Electronic mass enhancements, which stem from the $\Sigma'(\omega)$ (see Supplementary Text S6), provide a consistent picture from another perspective. Fig 3D shows that the mass enhancement of the nodal cut (cut 2) of $Pr_4Ni_3O_8$ is almost 5, where the various cuprate compounds in Fig. 4G show mass enhancements $\leq 2$. Here we only focused on the comparison to the strange metal state/normal state of the cuprate, which is the precursor state of high $T_C$ superconductivity (see Supplementary Text S8 for a more detailed discussion). The significantly larger mass enhancement in $Pr_4Ni_3O_8$ (over 2.5 times of that in cuprates) is consistent with the larger coupling constant $\lambda$ (or steeper slope) in $\Sigma''(\omega)$ (see Supplementary Text S6). The comparison to cuprates on both the imaginary self-energy and the mass enhancement indicates that the electronic correlation in $Pr_4Ni_3O_8$ is of a similar exotic form but significantly stronger than that in cuprates.

**Score Card and Discussion**

Our work provides the first direct observation of the electronic structure in the planar trilayer nickelate, where the Fermi pockets we found are similar to those found in the cuprates, consistent with its $d$ orbital occupancy with dominant $dx^2\text{-}y^2$ character. Importantly, the electron correlations found in $Pr_4Ni_3O_8$ are of a very similar exotic type as in cuprates but are somewhat stronger, as reflected by both the larger mass enhancement and linear slope in $\Sigma''(\omega)$.

With this as well as with other experimental data we can put together a type of scorecard of the major physical attributes of the two families of compounds, as shown in Table 1. Row 1 (presence of superconductivity) we have already discussed, though the maximum $T_c$ of the cuprates remains (for now at least) significantly higher. Row 2 shows the result of recent RIXS measurements indicating the similarity of the magnetic excitations to the 2+ valence cuprates, albeit with a near-neighbor exchange constant that is almost a factor of two smaller[36]. Rows 3-5 are the new results directly from our paper, indicating the great similarity between the electronic structure and correlations of 1+ valence nickelates compared to 2+ valence cuprates, with the most dramatic difference being the factor of two increased correlation strength of the nickelates. This increased correlation strength is generally consistent with the concept that the correlation gap of the parent of the nickelates is closer towards the Mott-Hubbard limit (row 6) while the gaps in the cuprates are more charge-transfer like [2], though otherwise the similarities between the electronic structure and the correlations are very clear. Finally, row 7 asks the question whether the superconductivity in the nickelates should likely have $d_{x2-y2}$ pairing symmetry, similar to the cuprates and unlike the vast majority of other superconductors. We argue that it likely should be – not just from the general similarity of the two compounds, but more specifically from the fact that both compounds exhibit a set of flat bands very near (within ~ 40 meV) $E_F$ at the $(\pi,0)$ points of the Brillouin zone. This very high density of states (an extended van Hove singularity [37]) near $E_F$ will likely make these states the "drivers" of the superconductivity while the lower density of states from the zone diagonal will likely make them the "passengers". We argue that it is therefore likely that a gap of $d_{x2-y2}$ symmetry, which derives most of its energy gain from pairing at the $(\pi,0)$ points and which is generally supported by antiferromagnetic spin fluctuations of the type shown on row 2, should also likely be favored in the nickelate superconductors.

Previous works in cuprates have found that the correlation effects in the strange metal phase play a key role for enhancing Cooper pairing [35,38]. Likewise, a recent study on ferropnictides also shows a direct connection between the slope of the scattering rate and $T_C$ [33]. In this context, our result raises the important question of whether such stronger strange metal correlations could support high temperature superconductivity in $Pr_4Ni_3O_8$ [35,38]. According to the recent DFT and X-ray absorption studies [8,19,20], the charge transfer energy between O-$2p$ and Ni-$3d$ in the planar trilayer nickelate is much smaller than that of the infinite layer nickelate superconductor, which places the planar trilayer nickelate closer to the cuprates. The hybridization of $p$-$d$ states in $Pr_4Ni_3O_8$ gives a substantial super-exchange coupling $J$ of ~69 meV, as recently revealed by resonant inelastic X-ray spectroscopy [39]. Based on the $t$-$J$ model, the predicted $T_C$ of the electron-doped planar trilayer nickelate [20] would be significantly higher than the maximum $T_C$ (15K) of the infinite-layer nickelate [4] and potentially on par with optimally doped cuprates. The strong electron correlation we find in $Pr_4Ni_3O_8$ places these studies on a firmer experimental footing.

In summary, the massive strange-metal correlations in $Pr_4Ni_3O_8$ provide a new platform to explore strange metal physics and its connection to high temperature superconductivity. More generally, this distinct character to the 'overdoped' regime of planar nickelates may serve to bridge from our well-established descriptions of cuprates to our nascent understanding of the infinite layer nickelate superconductors.

## Materials and Methods

**ARPES measurements.** ARPES experiments were carried out on in-situ-cleaved single crystal surfaces at the Stanford Synchrotron Radiation Lightsource (SSRL) beamline 5-2 and Diamond Lightsource beamline I05. The ARPES data shown in the main text are measured at SSRL beamline 5-2 with energy resolution of 20meV. Data in Supplementary Materials Fig. S2 are measured at Diamond beamline I05 with energy resolution of 5meV. All data shown in the paper were measured with the photon energy of either 55eV or 84 eV unless otherwise noted. All Fermi surface maps shown in the paper are integrated intensity over $E \pm 6$ meV.

**Single-crystal growth and transport measurement.** $Pr_4Ni_3O_8$ single crystals (1–2 mm$^2$ × 0.5 mm) were obtained by reducing specimens cleaved from boules of $Pr_4Ni_3O_{10}$ (flowing 4% $H_2$/Ar gas, 350 °C, five days). High pressure single-crystal growth of $Pr_4Ni_3O_{10}$ was performed in an optical-image floating zone furnace (HKZ-1, SciDre GmbH) with 140 bar $O_2$ [12]. The crystal structure was identified using X-ray diffraction [19]. Resistivity of $Pr_4Ni_3O_8$ single crystals (Fig. 3F) was measured on a Quantum Design PPMS in the temperature range of 2–300 K using a conventional four-probe method with contacts made with silver paint.

**DFT calculation.**

The first principle calculations of $La_4Ni_3O_8$ and $Pr_4Ni_3O_8$ (result shown in Supplementary Materials Fig. S5) was performed with the projector-augmented wave (PAW) method, as implemented in the Vienna Ab-initio Simulation Package (VASP). The plane-wave cutoff energy was taken as 400 eV. A full relaxation of the structure was carried out up to the breaking conditions that the total energy change between two electronic steps being less than 1E-5 eV, and all forces being smaller than 1E-3 eV/Å. The Perdew-Burke-Ernzerhof GGA was used to calculate the electronic structure, with a k-mesh of 25*25*9.

The $Pr_4Ni_3O_8$ DFT band structure at the Fermi level is dominated by the Pr $4f$ states, intersecting the Ni $3d$ bands (see Supplementary Materials Fig. S5). This is inconsistent with our experimental observations. If one includes the effects of electron correlation and applies different Coulomb repulsion values $U$ to different orbital states, one can purposely move the $4f$ states away from $E_f$, and achieve a $3d$ band structure with similar Fermi surface and bandwidth to that of $La_4Ni_3O_8$ [22,23]. However, in this work, we need a DFT bare band structure (no electron correlation) as a reference to extract the many-body information from the ARPES results, and the $s$, $p$, and $d$ bands of $Pr_4Ni_3O_8$ and $La_4Ni_3O_8$ are extremely similar. Thus, we choose to use the bare band structure of $La_4Ni_3O_8$ in this work.

**Acknowledgments**

We thank Drs. D. H. Lu and M. Hashimoto at SSRL beamline 5-2, and Drs. T. Kim, C. Cacho at Diamond beamline I05 for technical assistance on the ARPES measurements, and for help and valuable discussions. We thank Dr Y. Cao for the help in taking the LSCO and BSCCO data and Dr. Genda Gu for preparing the LSCO and BSCCO single crystals. Use of the Stanford Synchrotron Radiation Lightsource, SLAC National Accelerator Laboratory, is supported by the U.S. Department of Energy, Office of Science, Office of Basic Energy Sciences under Contract No. DE-AC02-76SF00515. This work was carried out with the support of the Diamond Light Source, beamline I05 (proposal SI17595). Experimental and theoretical work at the University of Colorado was supported by Gordon and Betty Moore Foundation (GBMF9458/Dessau). The ARPES studies were supported by the DOE under grant DE-FG02-03ER46066. The work in the Materials Science Division of Argonne National Laboratory (crystal growth and characterization) was supported by the U.S. Department of Energy, Office of Science, Basic Energy Sciences, Materials Science and Engineering Division. This work at Shandong University was supported by the Qilu Young Scholar Program of Shandong University, and the Taishan Scholar Program of Shandong Province. **Author Contributions:** H.L. led the ARPES measurement and analysis. P.H., A.G.L., X.Z. and K.G. helped with the ARPES measurement. J.Z., H.Z. and J.F.M. prepared the single crystal samples and measured the resistivity of the samples. P.H. and D.S.D carried out the density functional calculations. H.L. and D.S.D. did the majority of the paper writing, with contributions from all coauthors. D.S.D. directed the overall project. **Competing interests:** The authors declare no competing interests. **Data and materials availability**: All data needed to evaluate the conclusions in the paper are present in the paper and the Supplementary Materials.

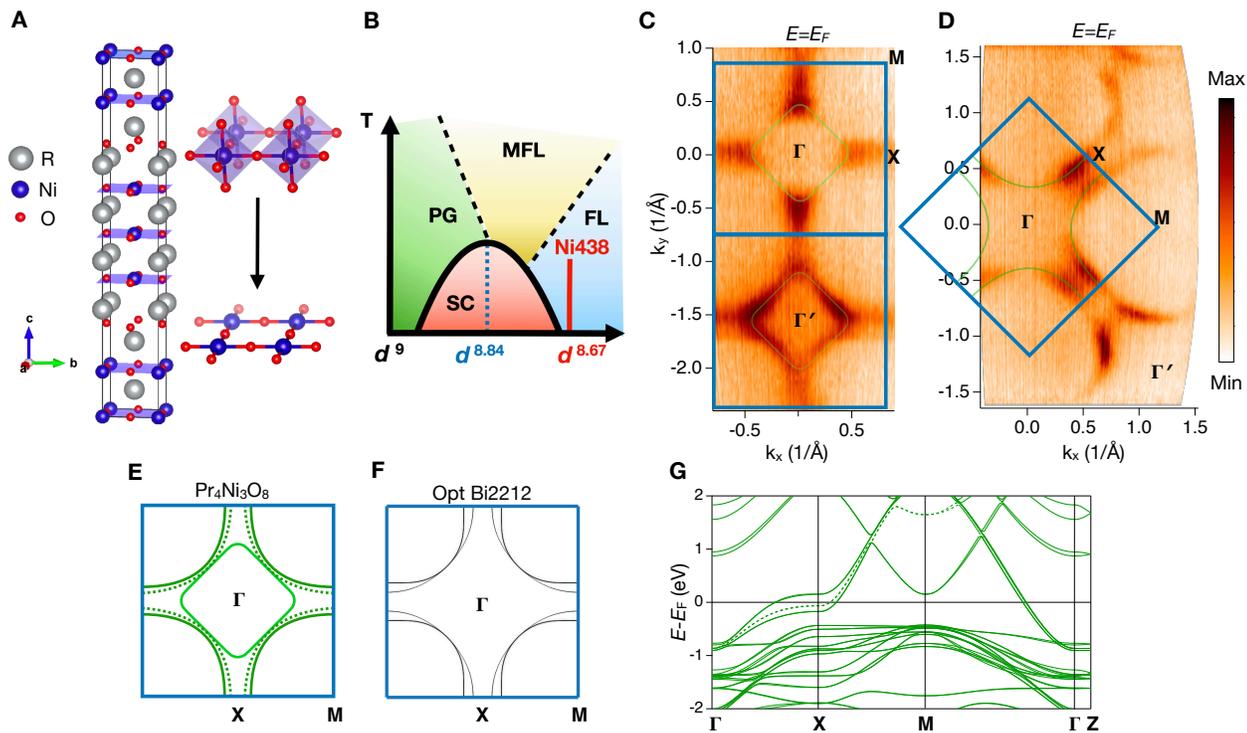

**Fig. 1. Crystal structure and Fermi surface of Pr$_4$Ni$_3$O$_8$.** (**A**) The unit cell of Pr$_4$Ni$_3$O$_8$ and a schematic showing the change from layered perovskite structure to square planar structure by the reduction of the apical oxygen. (**B**) Simplified doping phase diagram of cuprate superconductors, with the planar trilayer nickelate doping level (red line) residing in the overdoped region. (**C**) The Fermi surface maps taken with 84 eV photon energy that emphasizes the electron pocket around the Gamma point. (**D**) The Fermi surface map taken with 55 eV photons, which emphasizes the hole pocket centered around the zone corners (**E**) Schematics of the Fermi surface. The solid green pockets indicate the electron and hole pockets shown in panels (**C**) and (**D**). Two of the three Fermi surfaces are extremely similar to those of cuprates, including effective doping level. The third band in nickelate accommodates extra holes. The dashed hole pocket is an extra band shown in DFT result [see panel (**G**)]. (**F**) A simulated Fermi surface of an optimally doped Bi$_2$Sr$_2$CaCu$_2$O$_{8+x}$ (Bi2212) with the tight binding model parameters from ref [40]. (**G**) Calculated band structure using the GGA method. Here, we show the DFT result of La$_4$Ni$_3$O$_8$ instead of Pr$_4$Ni$_3$O$_8$ to avoid the issues related to the Pr-4$f$ states. The DFT results of Pr$_4$Ni$_3$O$_8$ are shown in Supplementary Materials Fig. S5, and the issue from the Pr-4$f$ is discussed in the method section and Supplementary Materials S7, and also in another DFT paper on this material [20].

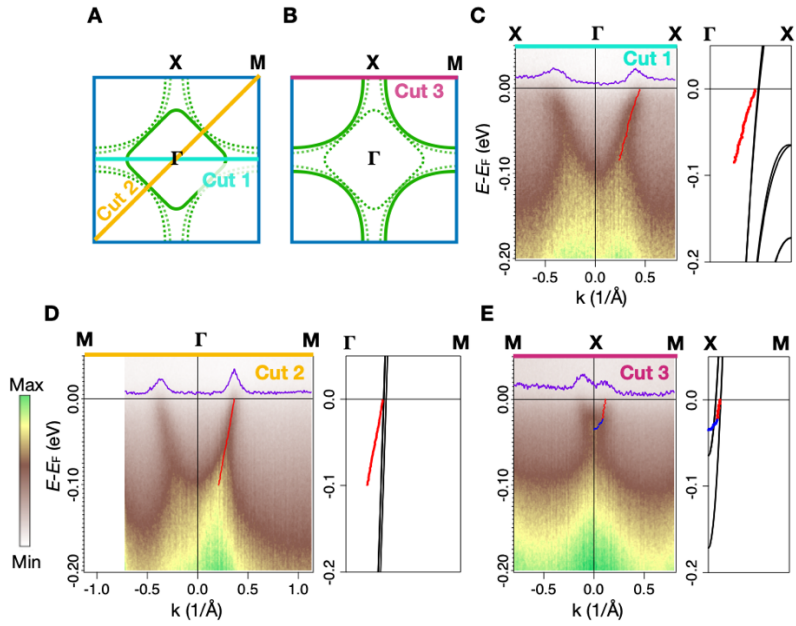

**Fig. 2. High symmetry cuts of different pieces of Fermi surface of Pr$_4$Ni$_3$O$_8$. (A-B)** Schematics of Fermi surface. We use 84eV photons to emphasize the electron pocket [panel (**A**)], and 55 eV photons to emphasize the hole pocket [panel (**B**)]. (**C-D**) High symmetry cuts taken with 84 eV photons and (**E**) 55 eV photons. All spectra in this figure are taken at temperature T=22K. The purple lines in panel C-D are the MDCs at the Fermi energy. Band dispersions extracted from the peak positions of MDCs (red dots) and EDCs (blue dots) are plotted in the spectra. To the right of each spectrum, we compare the DFT bands (black curves) of La$_4$Ni$_3$O$_8$ to the MDC and EDC dispersions extracted from the ARPES spectra.

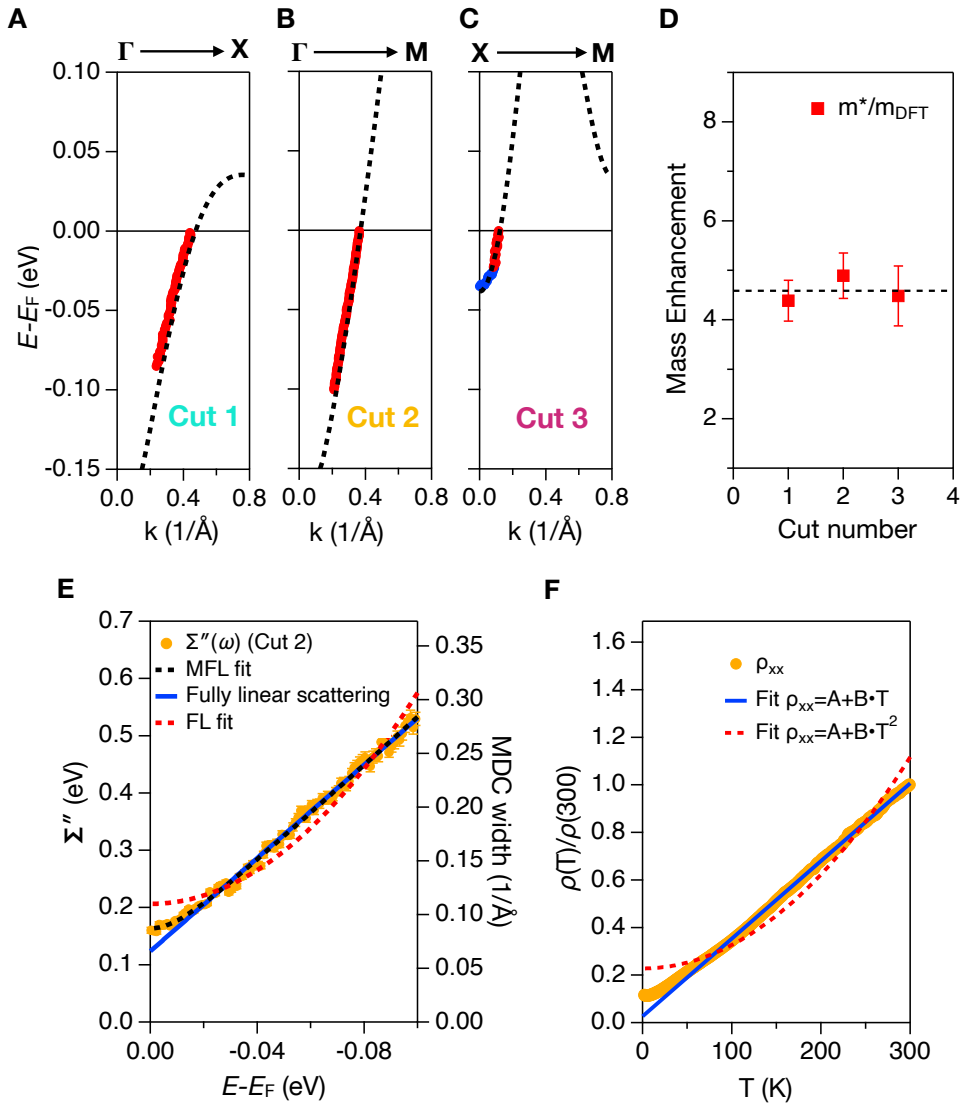

**Fig. 3. Large effective mass and linear-dependent quasiparticle scattering rate. (A-C)** MDC (red dots) and EDC (blue dots) dispersions extracted from ARPES spectra in Fig 2B-D and overlaid with the DFT bands renormalized to have a factor of 4-5 larger mass. **(D)** The mass enhancement extracted from the ARPES dispersion. The fact that the mass enhancement is approximately k-independent indicates that the correlation effects have a local origin, as expected for a Mott-type interaction. **(E)** The extracted imaginary part of the self-energy (inverse quasiparticle lifetime) from Cut 2. As the band disperses linearly near the Fermi level, the imaginary part of self-energy can be written as $\Sigma''=v_b\Gamma_{MDC}/2$, where $\Sigma''$ is the imaginary part of self-energy, $\Gamma_{MDC}$ is the FWHM of the MDCs, and $v_b$ is the bare band velocity obtained from the DFT band dispersion. The energy dependence of $\Sigma''(\omega)$ shows a good fit to the Marginal Fermi Liquid (MFL) model (black dashed curve). The standard deviation of the MDC widths from the fitting are included with the data points but are barely visible beyond the point size. **(F)** The resistivity vs. temperature. The dashed red curve is the fit with a quadratic function, the green solid curve is the linear fit. The resistivity data

are taken from ref [19]. The vertical bars in panel (**D**) represent the errors from extracting the band mass from the ARPES dispersions.

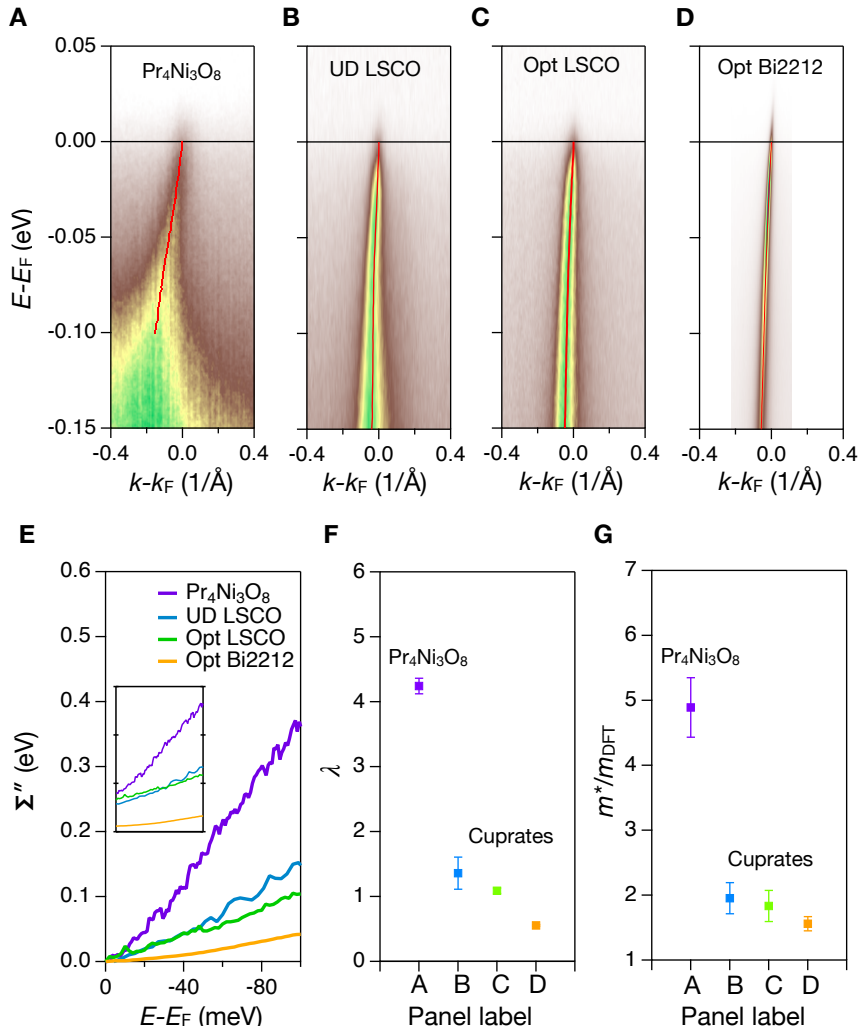

**Fig. 4. Comparing electronic correlation effects of Pr$_4$Ni$_3$O$_8$ and cuprates. (A-D)** ARPES spectra (cut 2, or nodal cut) of Pr$_4$Ni$_3$O$_8$ in comparison to cuprates. The cuprate samples are underdoped (UD, x=0.08) La$_{2-x}$Sr$_x$CuO$_2$ (LSCO), near optimally doped (Opt, x=0.17) LSCO, and optimally doped Bi$_2$Sr$_2$CaCu$_2$O$_{8+x}$ (Bi2212). All cuts are taken along (0,0)-($\pi$,$\pi$) or the nodal cuts. The red dots denote the MDC peak positions. (**E**) Imaginary self-energy ($\Sigma''$) with offset to the value at $E_F$. The non-offset plot is shown in the inset. (**F**) The slope parameter $\lambda$ extracted from the $\Sigma''(\omega)$ in panel E. The dimensionless coupling constant $\lambda$ is greater than 4 for the nickelate, whereas the cuprates ones are around 1. (**G**) The mass enhancement extracted from the band dispersions displayed in panel **A-D**, showing a 2 to 3-fold larger mass enhancement of nickelate compared to that of cuprates, which we argue is consistent with the increased slope ($\lambda$) of panel E. The LSCO data is taken at T=25K (UD) and T=35K (Opt), whereas the Opt Bi2212 data is taken at T=100K. All these temperatures are higher than the $T_C$ of the correspondent samples. The vertical bars in panel (**E**) and (**F**) represent the errors from extracting the slope parameter $\lambda$ from $\Sigma''(\omega)$ and the band mass from the ARPES dispersions, respectively.

| | 2+ valence cuprates | 1+ valence nickelates | Experimental observation (nickelates) |
|---|---|---|---|
| **1) Superconductivity** | to ~ 160K | to ~ 10K (for now) | D. Li et al.  (ref 4) |
| **2) Magnetic excitations** | Spin 1/2 in AFM square lattice. $J_1$ ~ 110 meV, $J_2$ ~ -11 meV | Spin 1/2 in AFM square lattice. $J_1$ ~ 64 meV, $J_2$ ~ -10 meV | H. Lu et al (ref 36) |
| **3) Fermi surface shape, geometry, etc.** | Extremely similar, including presence of vHs (flat bands) few tens of meV below $E_F$ at $(\pi,0)$ giving dominant 4-fold DOS contribution.  This should support d-wave electronic pairing. Additional band accomodates extra hole doping. | | This work. |
| **4) Electronic correlations - mass enhancement** | ~ factor of 2 relative to DFT. | ~ factor of 4 relative to DFT. Consistent with more Mott-like correlations. | This work. |
| **5) Electronic correlations – scattering rates.** | - Anomalously linear with E,T. Departure from Landau Fermi Liquid paradigm.  Termed "strange metal", "Marginal Fermi Liquid", etc. | - Anomalously linear with E,T. Departure from Landau Fermi paradigm.  Slope is ~ twice that in cuprates.  Consistent with factor of two in mass enhancements. | This work. |
| **6) Mott vs. Charge transfer gap.** | More charge-transfer like. | More Mott-like.  Consistent with stronger correlations. | This work, plus Goodge et al (ref 8). |
| **7) SC pairing symmetry** | $d_{x2-y2}$ | Also $d_{x2-y2}$ due to similarity in vHs at $(\pi,0)$? | To be determined. |

**Table 1.** Scorecard comparing the main physics topics between 2+ valence cuprates and 1+ valence nickelates.  For the widest generality we have mixed the "best" or cleanest results from infinite layer and trilayer nickelates.  The similarities between the two columns are overall quite dramatic, though with some numerical differences that may or not be important.